\documentclass[Physsubmission, Phys]{SciPost}

\hypersetup{
    colorlinks,
    linkcolor={red!50!black},
    citecolor={blue!50!black},
    urlcolor={blue!80!black}
}

\usepackage[bitstream-charter]{mathdesign}
\DeclareSymbolFont{usualmathcal}{OMS}{cmsy}{m}{n}
\DeclareSymbolFontAlphabet{\mathcal}{usualmathcal}

\begin{document}

\begin{center}{\Large \textbf{
Drell-Yan $p_{\bot}$ with NLO-matched Parton Branching TMDs at energies from fixed-target to LHC \\
}}\end{center}

\begin{center}
Aleksandra Lelek \textsuperscript{1$\star$}
\end{center}

\begin{center}
{\bf 1} University of Antwerp, Belgium

* aleksandra.lelek@uantwerpen.be
\end{center}

\begin{center}
\today
\end{center}


\definecolor{palegray}{gray}{0.95}
\begin{center}
\colorbox{palegray}{
  \begin{tabular}{rr}
  \begin{minipage}{0.1\textwidth}
    \includegraphics[width=22mm]{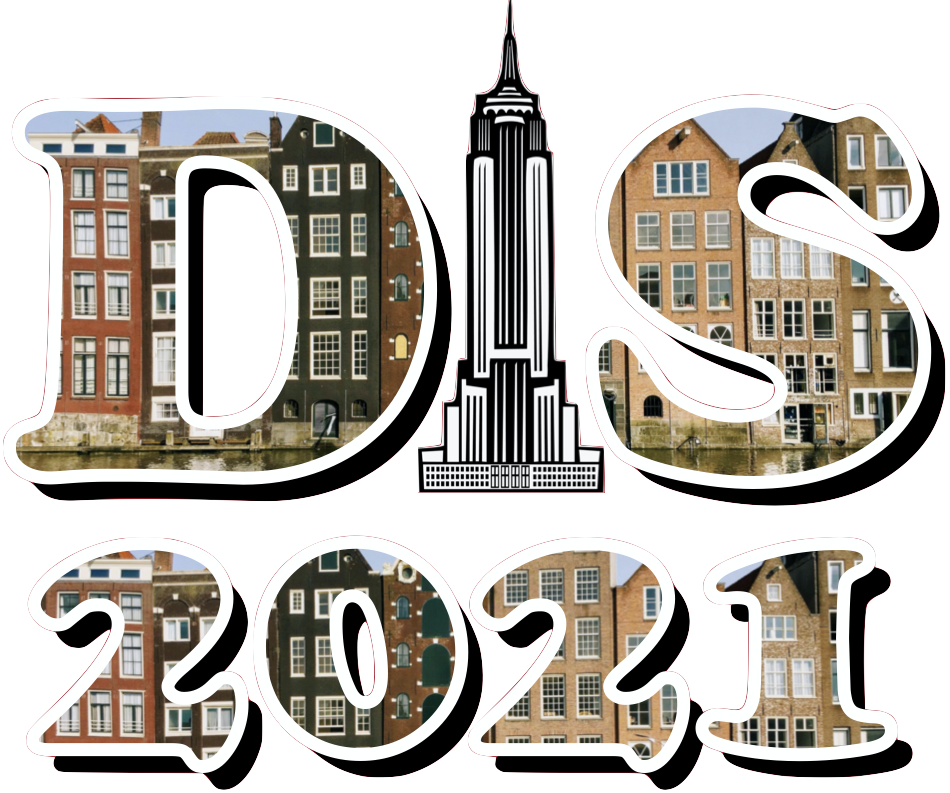}
  \end{minipage}
  &
  \begin{minipage}{0.75\textwidth}
    \begin{center}
    {\it Proceedings for the XXVIII International Workshop\\ on Deep-Inelastic Scattering and
Related Subjects,}\\
    {\it Stony Brook University, New York, USA, 12-16 April 2021} \\
    \doi{10.21468/SciPostPhysProc.?}\\
    \end{center}
  \end{minipage}
\end{tabular}
}
\end{center}

\section*{Abstract}
{\bf
The description of the Drell-Yan (DY) transverse momentum spectrum requires matching of fixed order QCD calculations with soft gluon resummation up to all orders in the QCD coupling. It has been noticed in the literature that a consistent description of DY data in a wide kinematic regime from fixed-target to LHC energies is problematic. In this talk the predictions for transverse momentum spectrum of DY data coming from experiments in very different kinematic ranges (NuSea, R209, Phenix, LHC 8 TeV and 13 TeV center-of-mass energies $\sqrt{s}$) are calculated by applying transverse momentum dependent (TMD) parton distributions obtained from the Parton Branching (PB) method, combined with the next-to-leading-order (NLO) calculation of the hard process in the MCatNLO method. We discuss the problems involved in matching of the fixed order calculation and resummation, especially in the moderate to low mass and $p_{\bot}$ region accessible at fixed target experiments. We find that at low DY mass and low $\sqrt{s}$ even in the region of $p_{\bot}/m_{DY}\sim 1$ the contribution of multiple soft gluon emissions (included in the PB-TMDs) is essential to describe the measurements, while at larger masses and LHC energies the contribution from soft gluons in the region of $p_{\bot}/m_{DY}\sim 1$ is small.
}


\section{Introduction}
\label{sec:intro}
The precise description of Drell-Yan (DY) lepton pair production is crucial for our understanding of many aspects of QCD such as evolution, factorization, resummation. DY data are used in parton distribution functions (PDFs) extraction. As a very clean production channel, DY is used as a standard candle in precision electro-weak measurements. Despite so many applications, the consistent description of DY data in a wide kinematic range is still problematic. Problems arise especially in DY observables with multiple energy scales involved such as the  DY $p_{\bot}$ spectrum. 
In different $p_{\bot}/Q$ regimes (where $Q$ is the invariant mass of the DY lepton pair) different physics dominates:
in the high $p_{\bot}$ region (i.e. 
$p_{\bot}
{\raisebox{-.6ex}{\rlap{$\,\sim\,$}} \raisebox{.4ex}{$\,>\,$}}
Q$) the $p_{\bot}$ spectrum is  expected to be described by fixed order QCD calculation within collinear factorization \cite{Collins:1989gx}. In the low $p_{\bot}$ region ($p_{\bot}<<Q$)
soft gluon radiation plays an important role and  logarithms of $p_{\bot}/Q$ need to be resummed up to all orders in QCD running coupling. Soft gluon resummation can be performed using  transverse momentum dependent (TMD) factorization formulas (such as analytical Collins-Soper-Sterman (CSS) approach \cite{Collins:1984kg} or high energy ($k_{\bot}$) factorization \cite{Catani:1990xk,Catani:1990eg}) or Parton Shower (PS) procedures within Monte Carlo (MC) generators. 
The methods used in the low and high $p_{\bot}$ regions have to be combined properly to obtain accurate description in the middle $p_{\bot}$ region. 
Recent study \cite{Bacchetta:2019tcu} showed that the  collinear perturbative fixed-order calculations are not able to describe the DY $p_{T}$ spectra for $p_{\bot}\sim Q $ at fixed target experiments. 
In this work we address this issue by using 
the Parton Branching (PB) approach \cite{Hautmann:2017xtx,Hautmann:2017fcj,BermudezMartinez:2018fsv}.

\section{The Parton Branching method}

Usually, predictions for production processes in QCD collider physics are obtained from collinear factorization which factorizes the cross section into collinear PDFs and partonic process. 
The collinear factorization works well for sufficiently inclusive, single scale observables. However, for processes with more scales involved soft gluons need to be resummed up to all orders in QCD running coupling with the methods mentioned above.  
The PB method  is an MC approach to obtain QCD predictions based on TMD PDFs, called also TMDs. The idea behind is to promote the collinear factorization into a TMD-dependent one: the cross section is written as a convolution of PDFs and partonic process 
where now  both  depend on the transverse momentum $k_{\bot}$ of the parton. 

The PB method can be divided in two main stages: first, it provides the TMD evolution equation \cite{Hautmann:2017fcj} from which the TMDs can be obtained.
This equation has a structure similar to  DGLAP:
it is  based on the unitarity picture where parton evolution is expressed in terms of resolvable branching probabilities, provided by the real emission DGLAP splitting functions and non-resolvable branching probabilities, given by  Sudakov form factors.
The initial distribution consists of collinear factor and a gaussian factor (in a simple parameterization) for intrinsic $k_{\bot}$ distribution. Then, the transverse momentum is calculated at each branching. The PB method uses angular ordering (AO) \cite{Hautmann:2019biw} in a  similar way to \cite{Marchesini:1987cf}: the  angles of the emitted partons increase from the hadron side towards hard scattering. With AO, soft gluon resummation is included properly.  Important property of the PB TMDs is that one can obtain collinear PDF (or integrated TMD, iTMD) by integrating the PB TMD over $k_{\bot}$. 
The parameters of the TMDs initial distributions  are fitted to HERA DIS data \cite{BermudezMartinez:2018fsv} with \texttt{xFitter} \cite{Alekhin:2014irh}.  The PB TMDs and iTMDs can be accessed via  TMDlib \cite{Abdulov:2021ivr}, a library collecting  TMDs from different approaches. The PB PDFs can be also used within LHAPDF. The second stage of the PB method 
is to obtain predictions with  PB TMDs for QCD collider observables by using the TMDs in TMD MC generators. PB TMDs have recently been implemented in  
the MC generator CASCADE \cite{Baranov:2021uol}. 
The recipe to use TMDs to obtain collider predictions was first proposed in \cite{BermudezMartinez:2018fsv} and further extended for next-to-leading order (NLO) in \cite{BermudezMartinez:2019anj} where PB TMDs were combined with NLO matrix element (ME) within the  MADGRAPH5$\_$AMC@NLO (denoted here as MCatNLO) 
approach\cite{Alwall:2014hca}. 
MCatNLO generates the collinear NLO ME using iTMD in LHAPDF format. Then, an extra operation is needed to transform collinear  ME into a $k_{\bot}$- dependent ME: $k_{\bot}$ is  added to the event record according to the TMD corresponding to the iTMD from which the ME was initially generated. 
In order to combine  NLO ME with parton showers (PS)  and to avoid possible double counting, the standard MCatNLO method uses subtraction terms for soft and collinear contributions. The role of PB TMDs is similar to PS which is the reason  why  subtraction terms have to be used to combine PB TMDs with MCatNLO calculations.
The subtraction terms depend on the PS algorithm.  The AO used in PB is similar to Herwig6 \cite{Corcella:2002jc} so MCatNLO with Herwig6 subtraction is used to combine PB TMDs with MCatNLO. 

\begin{figure}
\begin{minipage}{0.33\linewidth}
\centerline{\includegraphics[width=0.85\linewidth]{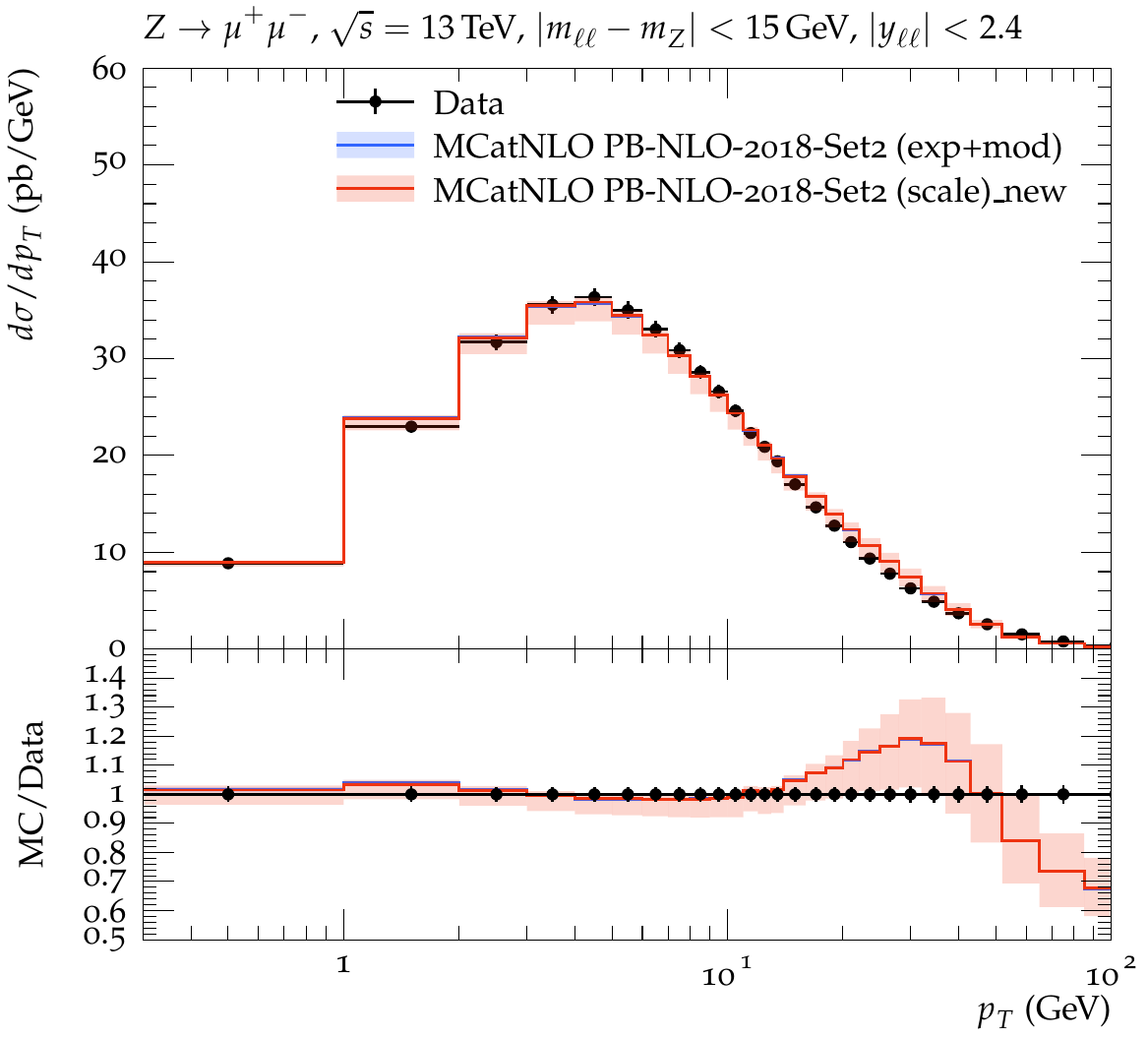}}
\end{minipage}
\hfill
\begin{minipage}{0.32\linewidth}
\centerline{\includegraphics[width=0.85\linewidth]{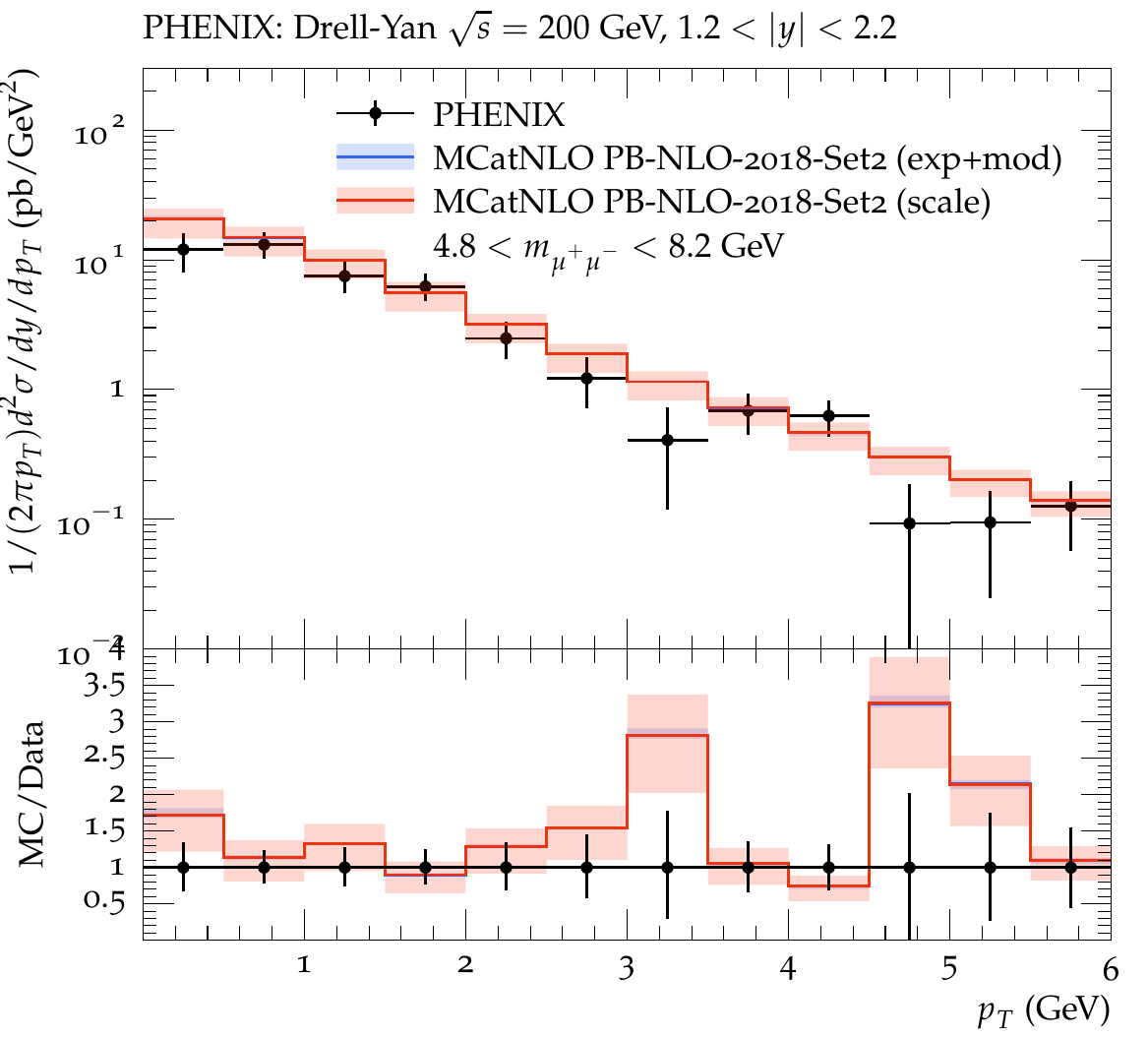}}
\end{minipage}
\hfill
\begin{minipage}{0.32\linewidth}
\centerline{\includegraphics[width=0.85\linewidth]{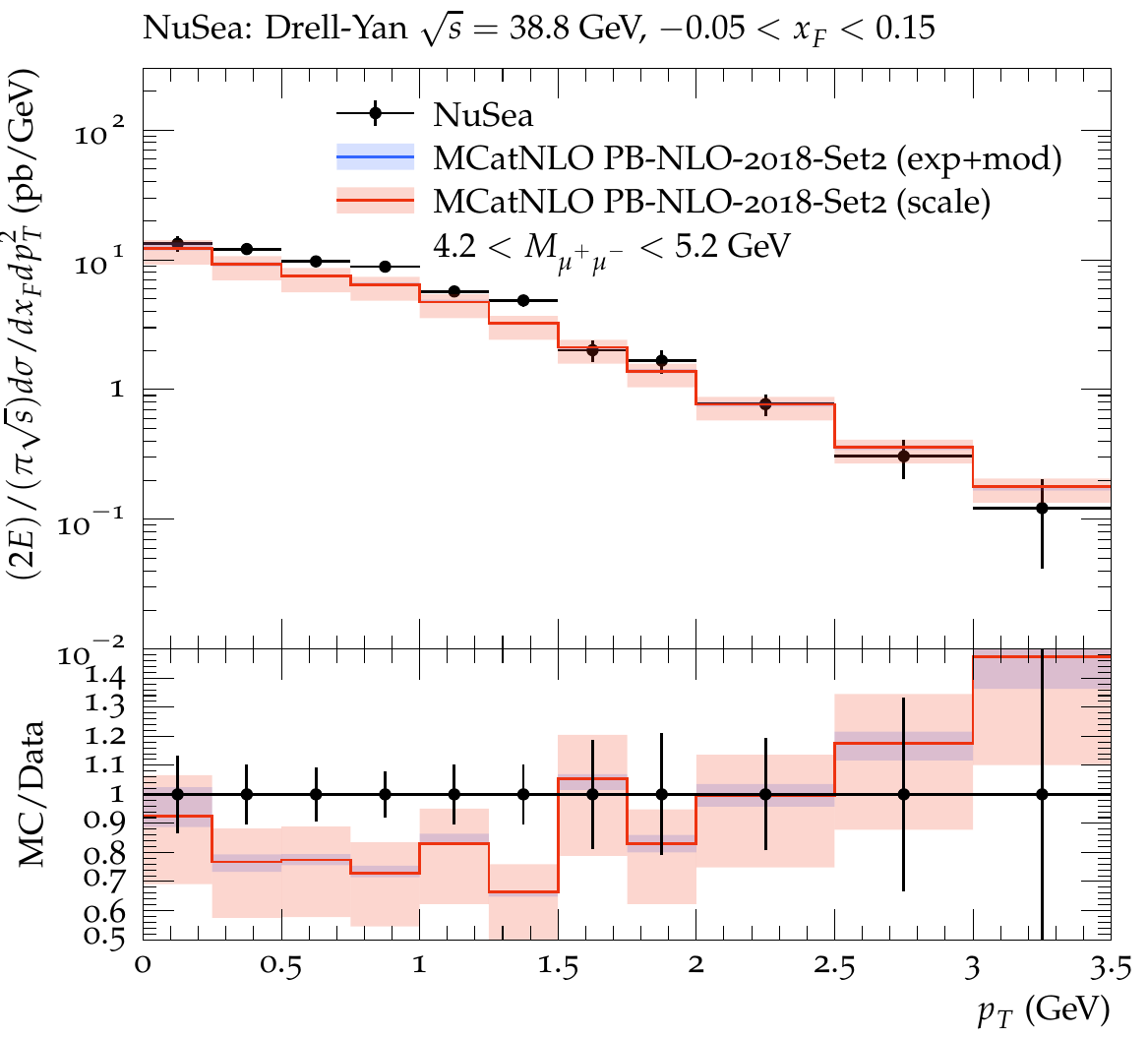}}
\end{minipage}
\caption[]{MCatNLO+PB TMD predictions for DY $p_{\bot}$ spectra  compared with CMS (left), PHENIX (middle) and NuSea (right) data  \cite{BermudezMartinez:2020tys}. }
\label{fig:predictions}
\end{figure}

\begin{figure}
\begin{minipage}{0.33\linewidth}
\centerline{\includegraphics[width=0.88\linewidth]{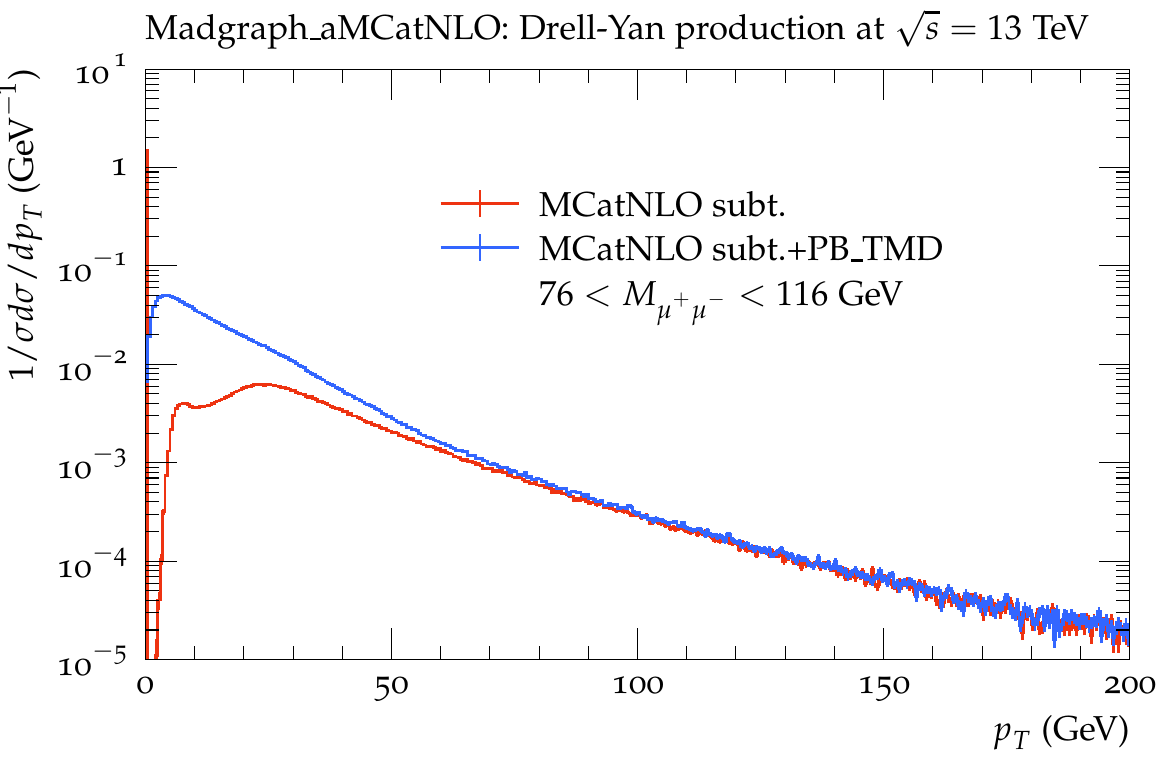}}
\end{minipage}
\hfill
\begin{minipage}{0.32\linewidth}
\centerline{\includegraphics[width=0.88\linewidth]{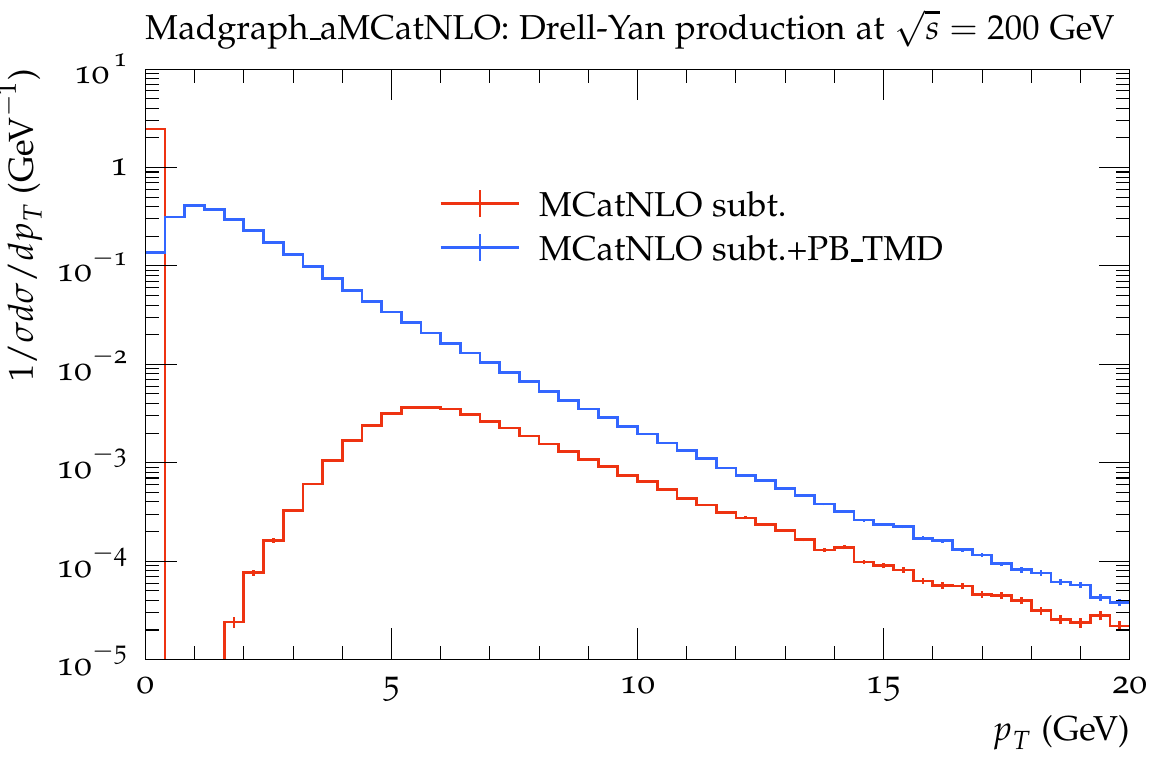}}
\end{minipage}
\hfill
\begin{minipage}{0.32\linewidth}
\centerline{\includegraphics[width=0.88\linewidth]{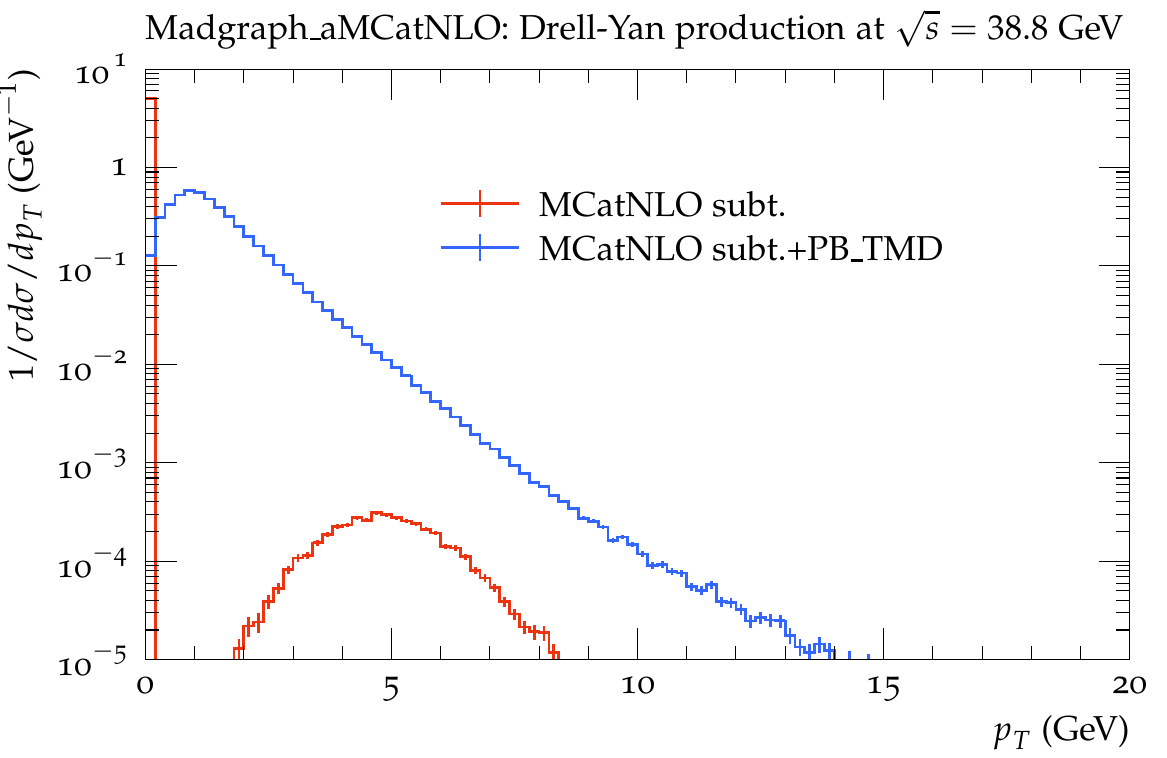}}
\end{minipage}
\caption[]{MCatNLO calculation with subtraction term (red) and full MCatNLO+PB TMD calculation (blue) at $\sqrt{s}$ corresponding to Fig.~\ref{fig:predictions} \cite{BermudezMartinez:2020tys}. }
\label{fig:subtraction}
\end{figure}

\section{Results for DY $p_{\bot}$}
The method described above was applied to DY data from different experiments at different  center of mass energies $\sqrt{s}$ and DY masses \cite{BermudezMartinez:2020tys}: NuSea \cite{Webb:2003bj}, R209 \cite{Antreasyan:1981eg}, PHENIX \cite{PHENIX:2018dwt}, ATLAS  \cite{ATLAS:2015iiu}  and CMS \cite{CMS:2019raw}. The predictions for DY $p_{\bot}$ spectra coming from CMS, PHENIX and NuSea obtained with  PB-NLO-HERAI+II-2018-set2 TMD PDF~\cite{BermudezMartinez:2018fsv}+MCatNLO are shown in Fig.~\ref{fig:predictions}. A good description is obtained in all these kinematic regimes in small and middle $p_{\bot}$ range. To obtain a proper prediction in the high $p_{\bot}$, higher jet multiplicities have to be taken into account what was recently achieved in \cite{Martinez:2021chk}. Important to stress is that  once the parameters of the initial distributions are fitted, there are no other parameters which require adjustment and all those predictions were obtained with the same settings. 
In Fig.~\ref{fig:subtraction} the MCatNLO+PB TMD prediction (blue) is compared to subtracted ME calculation (red) for $\sqrt{s}$ corresponding to plots in Fig.~\ref{fig:predictions}. At low DY mass and
low $\sqrt{s}$,  even in the region of $p_{\bot} \sim Q$, the contribution of 
soft gluon emissions contained in PB TMDs is essential to describe the
data. The situation at LHC energies and larger masses is different: 
here  the contribution from soft
gluons in the region of $p_{\bot} \sim Q$ is small and the spectrum is dominated by hard real emission.

\subsection{Remark on intrinsic $k_{\bot}$ distribution}
In the current fit procedure within \texttt{xFitter}, the width of the intrinsic $k_{\bot}$ gaussian  
  of the PB TMDs is not constrained  because of  low sensitivity of the HERA data to intrinsic $k_{\bot}$ and it is fixed to $\sigma^2=q_s^2/2$ with $q_s=0.5 \; \textrm{GeV}$. However, it was shown in \cite{BermudezMartinez:2020tys} that the fixed target DY data are sensitive to the intrinsic $k_{\bot}$ and are described best  with  $q_s \in (0.2,0.4)\;\rm{GeV}$. This is  close to the value initially chosen in~\cite{BermudezMartinez:2018fsv}.

\section{Conclusion}
The description of DY $p_{\bot}$ requires different methods in different $p_{\bot}$ regimes and its precision depends on matching between them. 
In this work we presented predictions for DY $p_{\bot}$  at different $\sqrt{s}$ and for different DY masses obtained with PB TMDs combined with NLO ME via  MCatNLO method. 
We confirm the result of \cite{Bacchetta:2019tcu}  that   collinear fixed-order calculations are not able to describe the DY $p_{T}$ for $p_{\bot}\sim Q $ of order few GeV 
at fixed target experiments. We notice that  soft gluon contributions have to be included to describe the data in this regime. This is different from the LHC,  where for $p_{\bot}\sim Q $ the DY $p_{\bot}$   is dominated by the real hard emission.

\section*{Acknowledgements}
The presented results were obtained in collaboration with  A. Bermudez Martinez, P. Connor, D. Dominguez Damiani, L. I. Estevez Banos, F. Hautmann, H. Jung, J. Lidrych, M. Mendizabal Morentin, M. Schmitz, S. Taheri Monfared, Q. Wang, T. Wening, H. Yang and R. Zlebcik.
\paragraph{Funding information}
A. L. acknowledges funding by Research Foundation-Flanders (1272421N).

\bibliography{OlaLelek.bib}

\end{document}